\documentclass[journal]{IEEEtran}
\usepackage{amsmath, amsfonts, amssymb}
\usepackage{mathtools}
\usepackage{bm}
\usepackage{graphicx}
\usepackage{cite}
\usepackage{algorithm}
\usepackage{algorithmic}
\usepackage{booktabs}
\usepackage{multirow}
\usepackage{hyperref}
\hypersetup{hidelinks}
\usepackage{array}
\usepackage{kvdefinekeys}
\usepackage{booktabs}
\usepackage{multirow}
\usepackage{xcolor}
\usepackage{enumitem}
\usepackage{threeparttable}

\newcommand{\norm}[1]{\left\lVert #1 \right\rVert}

\begin{document}

\title{AC-Informed DC Optimal Transmission\\ Switching  via Admittance Sensitivity-Augmented\\ Constraints and Repair Costs}
\author{
  Rahul~K. Gupta,~\IEEEmembership{Member,~IEEE} 
  \thanks{The author is with the School of Electrical and Computer Science, Washington State University (e-mail: \texttt{rahul.k.gupta@wsu.edu}).}
}
\maketitle
\begin{abstract}
AC optimal transmission switching (AC-OTS) is a computationally challenging problem due to the nonconvexity and nonlinearity of AC power-flow (PF) equations coupled with a large number of binary variables. A computationally efficient alternative is the DC-OTS model, which uses the DC PF equations, but it can yield infeasible or suboptimal switching decisions when evaluated under the full AC optimal power flow (AC-OPF). To tackle this issue, we propose an AC‑Informed DC Optimal Transmission Switching (AIDC-OTS) scheme that enhances the DC-OTS model by leveraging first- and second-order admittance sensitivities-based constraints and repair/penalty costs that guide the DC-OTS towards AC-feasible topologies. 
The resulting model initially is a Mixed-Integer Quadratically Constrained Quadratic Program (MIQCQP), which we further reformulate into solver-friendly representations, such as a Mixed-Integer Second-Order Cone Program (MISOCP) and a Mixed-Integer Linear Program (MILP). This proposed scheme yields switching topologies that are AC-feasible, while maintaining computational tractability. We validate the proposed scheme using extensive simulations across a large set of PGlib test cases, demonstrating its effectiveness, with performance benchmarks against original DC-OTS and other OTS formulations such as LPAC-OTS and QC-OTS. 
\end{abstract}

\section{Introduction}
\label{sec:intro}
\subsection{Background and Literature Review}
Transmission line switching has emerged as a flexible operational strategy for transmission networks, enabling operators to optimize grid topology by switching some lines on or off, a scheme known as optimal transmission switching (OTS) \cite{o2005dispatchable, fisher2008optimal}. OTS has been applied to achieve a wide range of operational objectives, including minimizing generation dispatch cost \cite{hedman2008optimal}, alleviating line congestion \cite{khanabadi2011transmission, salkuti2018congestion}, reducing load shedding \cite{amraee2017controlled}, and enhancing system reliability, etc. More recently, OTS has been explored as a tool for mitigating wildfire-related risks \cite{ harris2025integrated, huang2025machine}, where the objective is to minimize load shedding associated with wildfire‑affected transmission assets. In planning contexts, OTS has also been used as a strategy to defer capital investments when integrated with transmission expansion studies \cite{khodaei2010transmission, yuan2022resilience, piansky2025optimizing}.

As the OTS problem involves closing/opening transmission lines that are modeled by binary variables, it is often computationally challenging to solve in its original form. 

The OTS problem with true nonlinear AC power-flow equations, referred to as the AC‑OTS problem, is a mixed‑integer nonlinear program (MINLP) and does not scale well for large test cases. One way to tackle this issue is by approximating the AC power flow equations with DC power flow equations, resulting in the DC optimal transmission switching (DC‑OTS) problem. DC‑OTS uses the DC power flow approximation, which assumes unity voltage magnitudes and ignores losses and reactive power injections \cite{stott2009dc}. With this simplification, the OTS problem becomes a mixed‑integer linear program (MILP). Although DC‑OTS is efficient to solve thanks to the linearization provided by the DC approximation, it is often infeasible with respect to the AC nonlinear equations. It may also yield unrealistically low operating costs that cannot be achieved under the true AC power flow constraints \cite{barrows2014correcting}.

Several recent works have proposed algorithms to improve upon DC-OTS and achieve performance closer to the AC‑OTS problem. One widely used approach is to deploy various convex relaxation schemes that were originally applied to the AC‑OPF problem and later extended to the AC‑OTS problem. For example, a quadratic‑constrained OTS (QC‑OTS) formulation has been proposed in \cite{hijazi2017convex, coffrin2015_qc}, which relies on a quadratic relaxation of the AC power flow equations. Although these schemes provide a lower bound on the operational cost of the original AC‑OTS problem, they may still be infeasible under the AC equations. In \cite{guo2025tightening}, the QC‑OTS model is further refined with stronger constraints, offering tighter approximations to the AC‑OTS. Another approach uses the second‑order‑cone‑program (SOCP) relaxation of the AC‑OPF within the OTS problem, leading to an MISOCP formulation \cite{kocuk2017new}. The MISOCP formulation is more tractable than QC‑OTS, though it can be less accurate. A different scheme is based on the linearization of the AC power flow equations (LPAC) \cite{coffrin2014linear}, which is computationally efficient compared to nonlinear relaxations, such as QC and SOCP formulations. The LPAC‑based OTS problem, formulated in \cite{coffrin2014linear}, more often yields AC‑feasible solutions compared to DC‑OTS.

Despite these advances of the above relaxations and linearization schemes, a key disadvantage is that they increase computational complexity, even though the resulting solutions are better than those from the DC‑OTS problem. As highlighted in recent works, AC‑relaxation‑based OTS may still produce suboptimal topologies and can remain infeasible with respect to the true AC power flow equations, especially under changing topologies due to line switching. Consequently, there has been active research aimed at improving the computational efficiency of the AC‑OTS problem while obtaining AC‑feasible and near‑optimal solutions.

Another class of OTS schemes is the AC‑informed DC‑OTS approach, where the DC‑OTS problem is modified to produce solutions closer to the AC‑OTS problem. For example, the work in \cite{taheri2025ac} developed a scheme in which the network parameters are adjusted so that the DC‑OTS solutions better approximate the AC‑OTS solutions. The scheme relies on modifying network parameters based on offline training. It is proposed as a bilevel formulation, where the lower level solves the AC‑OPF problem, and the upper level optimizes the DC‑OPF network parameters to match the AC‑OPF solutions. The method incorporates a loss function representing branch currents, which makes it suitable for the OTS problem, as OTS decisions strongly depend on network congestion patterns. Although this scheme demonstrates improved AC‑feasible and near‑optimal solutions compared to standard DC‑OTS, it requires training, and its performance depends heavily on the quality and representativeness of the training data. Moreover, the training procedure itself requires repeatedly solving the original AC‑OPF, which is nonlinear, nonconvex, and computationally expensive.
\subsection{Proposed formulation and Contributions}
To tackle the above mentioned issues, in this paper, we propose a new approach for AC-informed DC-OTS (AIDC-OTS). We propose augmenting the DC-OTS objective with an AC-informed penalty/repair cost as well as constraints that enhance the AC feasibility and suboptimality of the original DC-OTS formulation. We derive a penalty objective using first- and second-order admittance sensitivities (i.e., first- and second-order derivatives of the network states, with respect to the admittance parameters), which is also minimized in the DC-OTS objectives. This objective term penalizes the topologies that may cause a large perturbation in the network's states (voltage magnitudes and angles), a principle that is usually followed by the network operators. Aside from the AC-repair penalty, it also includes a set of linear and quadratic constraints that are bounded by pre-computed tolerances as the maximum changes in the system states that may be allowed during the topology change. These tolerances are derived offline based on the base network admittance parameters and AC-OPF solutions. Combining the penalty objective and the constraints helps improve the AC-feasible solution as well as better AC-optimal solutions, when the optimized topology is evaluated with AC-OPF. 

The key contributions of the work are listed as
\begin{itemize}
  \item An \textbf{AC-informed DC-OTS formulation}: We augment the DC-OTS formulation with first and second-order admittance sensitivities informing the DC-OTS problem on the sensitivity of the line switching on the AC voltage magnitudes and angles. This is achieved by introducing (i) AC penalty cost and (ii) sensitivity-augmented constraints. The first-order admittance sensitivity terms help in minimal AC correction after a switch, whereas the second-order terms for curvature control remain in the local AC-validity region.
  \item \textbf{SOCP and Linear approximations}: The proposed scheme is MIQCQP, so we reformulate to obtain MISOCP and MILP reformulations.
  \item \textbf{Derivation of Second-order Admittance Sensitivities}: We derive the second-order admittance sensitivities, which are required in the sensitivity-based AC repair cost and constraints. 
    \item \textbf{Numerical Validation:} We validate the proposed scheme on several PGLib cases and compare the performance on \emph{AC-OPF-evaluated}  costs/violations against original DC-OTS, LPAC-OTS, and QC-OTS formulations.
\end{itemize}
\section{Problem Statement and DC-OTS Model}
We consider the problem of optimizing the topology of a transmission network with the objective to minimize the generator dispatch costs. In the following, first we review the DC-OTS model, then motivate the proposed scheme of the AC-informed DC-OTS model.
\subsection{DC Optimal Transmission Switching}
Let the transmission system consists of $N$ nodes, contained in set $\mathcal{N} = \{1, \dots, N\}$, $L$ lines contained in set $\mathcal{L} = \{1, \dots, L\}$. Let the generator indices are contained in $\mathcal{G} = \{1, \dots, G\}$, where $G$ represent number of generators. The line admittances of the network are denoted by $y_l = g_\ell + jb_\ell$, where $g_\ell$ and $b_\ell$ are conductance and susceptance. Let $z_\ell = \{0,1\}$ denotes binary variable for switching of line $\ell$, $P_g$ represent the variable for generator setpoint, $P_g^\text{min}, P_g^\text{max}$ being the generator setpoint limits,  $P_{d,i}$ refer to demand at node $i$, $\theta_i$ refer to the voltage angle at node $i$. Let the symbol $f_\ell$ refer to the flow in the line $\ell$ and $\bar{f}_\ell$ denote the flow limit.
The DC-OTS problem is formulated as
\begin{subequations}\label{eq:dc-ots}
\begin{align}
\min_{P_g,\bm{\theta},\bm{f},\bm{z}} \quad & \sum_{g\in\mathcal{G}} C_g(P_g) \label{eq:dc-ots-obj}\\
\text{s.t.}\quad 
& \sum_{g\in\mathcal{G}_i} P_g - P_{d,i} = \sum_{\ell\in\delta(i)} \sigma_{i\ell} f_\ell,\quad \forall i\in\mathcal{N}, \label{eq:balance_dc}\\
& f_\ell - b_\ell(\theta_i-\theta_j) \le M_\ell(1-z_\ell),\ \forall \ell\in\mathcal{L}, \label{eq:bigM1_dc}\\
& f_\ell - b_\ell(\theta_i-\theta_j) \ge -M_\ell(1-z_\ell),\ \forall \ell\in\mathcal{L}, \label{eq:bigM2_dc}\\
& -\bar f_\ell z_\ell \le f_\ell \le \bar f_\ell z_\ell,\quad \forall \ell\in\mathcal{L}, \label{eq:flowlimit_dc}\\
& P_g^{\min} \le P_g \le P_g^{\max},\quad \forall g\in\mathcal{G}, \label{eq:genlimit_dc}\\
& z_\ell\in\{0,1\},\quad \forall \ell\in\mathcal{L}. \label{eq:binary}
\end{align}
\end{subequations}
where $C_g(P_g)$ in \eqref{eq:dc-ots-obj} refers to the generator cost function, which is expressed as linear or quadratic cost depending on the test case. Eq.~\eqref{eq:balance_dc} refers to the power balance equation, where $\sigma_{il} \in \{-1, 0, 1\}$ represents the connection between the line $\ell$ and node index $i$. The symbol $\delta(i)$ refers to indices of all the lines connected at node $i$. The constraints in \eqref{eq:bigM1_dc}, \eqref{eq:bigM2_dc}, \eqref{eq:flowlimit_dc} express the power-flow limits on the lines using the Big-M method. The power flow in the line is expressed using the DC linear approximation $f_l = b_l(\theta_i - \theta_j)$.
The constant $M$ is chosen according to the scheme in \cite{moulin2010transmission}. The constraint in \eqref{eq:genlimit_dc} expresses the limits on the generator.  

As can be seen, the DC-OTS problem is an MILP, thanks to the DC power flow constraints, and it can be efficiently solved by existing open-source and commercial solvers. 

\subsection{AC-Infeasibility of DC-Optimal Topologies}\label{sec:motivation}
As widely reported in the literature \cite{barrows2014correcting, pineda2024learning, taheri2025ac}, the DC-OTS scheme may yield network topologies that are infeasible when evaluated under full AC power flow constraints, as well as with AC-OPF. This problem arises from simplifications inherent in the DC power flow approximation. These approximations include fixed unit voltage magnitudes at all buses, neglecting reactive power balance, and enforcing the line flow limits based on the active power. Such approximations are reflected in constraints in \eqref{eq:balance_dc}-\eqref{eq:genlimit_dc}.

Consequently, DC-OTS formulations do not capture voltage magnitude constraints, generator reactive power capability limits, or apparent power thermal limits. After line switching, however, the redistribution of reactive power flows and the removal of line charging susceptance can significantly alter voltage profiles and reactive power requirements. As a result, DC-feasible switching decisions may violate voltage magnitude constraints or exceed generator reactive power limits when evaluated on AC-OPF, rendering the topology infeasible.
Furthermore, DC-OTS ignores network losses and assumes an exact active power balance between generation and demand. In contrast, AC power flow requires additional generation to supply nonlinear real power losses, which may increase following topology changes that lengthen electrical distances or reduce network redundancy. A DC-optimal topology may therefore lack sufficient generation to satisfy AC power balance equations. 

In summary, DC-OTS feasibility does not guarantee AC feasibility because the DC model omits reactive power dynamics, voltage constraints, losses, and nonlinear network strength considerations that fundamentally govern post-switching AC behavior. This structural mismatch motivates the incorporation of AC feasibility awareness into DC-based OTS formulations, as pursued in this work.
\section{Proposed AC-Informed DC-OTS Scheme}
Despite the computational advantages of the DC-OTS model, it can be AC-infeasible and sub-optimal as noted in the previous section. Therefore, the objective is to propose a scheme that can enhance the feasibility as well as optimality, while benefiting from the scalability of the DC-OTS model.

In the following, we introduce a new optimization model referred to as AC-informed DC-OTS problem (AIDC-OTS), which augments the DC-OTS model with extra objectives and constraints using the first- and second- order admittance sensitivities. The first- and second- order admittance sensitivities are defined as derivatives of the network states with respect to the admittance parameters, and are used to inform the DC-OTS model of the impact of switching a line on the network states.

\subsection{Admittance Sensitivity-Informed Optimization Model to Improve AC Feasibility}
\label{subsec:AIDOTS:theory}
Optimizing network topology requires switching transmission lines in or out of service, actions that directly perturb the complex admittance matrix of the grid. Even small changes in this matrix can propagate nonlocally through the system, altering voltage magnitudes, phase angles, reactive power flows, and losses, which the DC power‑flow approximation cannot capture. As a result, relying solely on the DC model may lead to switching selections that perform well in the simplified setting but violate feasibility under the full AC physics.
To mitigate this gap, we aim to guide the switching decisions within the DC‑OTS formulation so that they remain consistent with AC power‑flow behavior. 

A key operational insight is that switching actions should avoid inducing large perturbations in voltage magnitudes or phase angles. To embed this intuition into the DC‑OTS framework, we propose explicitly modeling \textit{how each switching action influences the underlying AC network states.} This is achieved by leveraging the derivatives of the AC states, i.e., voltage magnitudes and phase angles with respect to line‑admittance parameters. We refer to these quantities as \emph{admittance sensitivities}.
We incorporate these sensitivities into the DC‑OTS formulation through a combination of \emph{penalty terms} and \emph{additional constraints}, enabling the model to better anticipate the AC feasibility impacts of topological changes. In particular, we introduce:
\begin{enumerate}
    \item \textbf{First‑order sensitivity terms}, which capture the linear (i.e., first‑order) changes in network states resulting from adjustments to the line‑switching decisions.
    \item \textbf{Second‑order sensitivity terms}, which account for the higher‑order effects of switching actions on the network states, thereby improving the approximation of AC behavior under topology variations.
\end{enumerate}
The construction of these two terms is detailed below.
\subsubsection{First-order admittance sensitivity based penalty and constraints}
Let's start with the power-flow equations, which express the power injection at bus $i$ in terms of the bus voltages and the network admittance matrix:
    $s_i = p_i + jq_i = |v_i| \sum_{k=1}^{n} Y_{ik} v_k^*.$
where $Y_{ik} =  G_{ik} + jB_{ik}$ are the elements of the admittance matrix $\mathbf{Y} \in \mathbb{C}^{N\times N}$ with real part $\mathbf{G}$ and imaginary part $\mathbf{B}$. We also define longitudinal admittance matrix as $\mathbf{Y}_{\mathcal{L}} = \texttt{diag}(\mathbf{y}) \in \mathbb{C}^{L\times L}$, such that $\mathbf{y} = \mathbf{g} + j\mathbf{b}$ are the vectors of line admittances. $\mathbf{Y} = \mathbf{A}^\top \mathbf{Y} \mathbf{A}$, where $\mathbf{A}$ is the incidence matrix of the network. 
Separating real and imaginary components yields the nonlinear active and reactive power equations as
\begin{subequations}
   \begin{align}
    p_i &=  |v_i| \sum_{k=1}^{N}|v_k| |Y_{ik}| \cos(\theta_i - \theta_k - \phi_{ik}) \label{eq:polar_P} \\
    q_i &=  |v_i| \sum_{k=1}^{N} |v_k| |Y_{ik}| \sin(\theta_i - \theta_k - \phi_{ik}) \label{eq:polar_Q}
    \end{align} 
\end{subequations}
where $|v_i|$ denotes voltage magnitude, $\theta_i$ is the voltage angle, and $|Y_{ik}|\angle\phi_{ik}$ is the complex equivalent admittance of the line between buses $i$ and $k$. 

Let the power-flow (PF) equations in \eqref{eq:polar_P}-\eqref{eq:polar_Q} be written in the following form as
\begin{subequations}
\begin{align}
    \mathbf{f}(\mathbf{x}, \boldsymbol{\alpha}) =
\begin{bmatrix}
\mathbf{p}(\mathbf{x}, \boldsymbol{\alpha}) \\
\mathbf{q}(\mathbf{x}, \boldsymbol{\alpha})
\end{bmatrix}
= \mathbf{0}
\label{eq:pf_eq}
\end{align}
where the state vector is
$\mathbf{x} =
\begin{bmatrix}
|\mathbf{v}| \\
\boldsymbol{\theta}
\end{bmatrix}$
and the parameter vector is
$\boldsymbol{\alpha} = \begin{bmatrix}
\Re (\mathbf{y}) \\
\Im (\mathbf{y})
\end{bmatrix} = 
\begin{bmatrix}
\mathbf{g} \\
\mathbf{b}
\end{bmatrix}$.
Lets say, switching a line $\ell$ alters the vector of longitudinal admittances from $\mathbf{y}$ to $\mathbf{y}'$ by $\Delta \mathbf{y}$, which can be expressed as 
\begin{align}
\mathbf{y}' = \mathbf{y} + (1-z_\ell)\Delta \mathbf{y}.
\end{align}

A first-order linearization of the PF equation in \eqref{eq:pf_eq}, gives
\begin{align}
    & \frac{\partial \mathbf{f}}{\partial \mathbf{x}}\Delta \mathbf{x} + \frac{\partial \mathbf{f}}{\partial \boldsymbol{\alpha}}\Delta \boldsymbol{\alpha} \approx 0 \\
    \text{Or,}~ & \frac{\partial \mathbf{f}}{\partial \mathbf{x}}\Delta \mathbf{x} + \Big[ \frac{\partial \mathbf{f}}{\partial \mathbf{g}}\Delta \mathbf{g} + \frac{\partial \mathbf{f}}{\partial \mathbf{b}} \Delta \mathbf{b}\Big]\approx 0 \label{eq:FOC_1}
\end{align}

Defining $\mathbf{J} = \dfrac{\partial \mathbf{f}}{\partial \mathbf{x}}$, $\mathbf{J}_\alpha = \Big[ \dfrac{\partial \mathbf{f}}{\partial \mathbf{g}}~\dfrac{\partial \mathbf{f}}{\partial \mathbf{b}} \Big]$, $\bm \alpha = [\Delta \mathbf{g}~\Delta \mathbf{b}]^\top$.  Eq.~\eqref{eq:FOC_1} can be rewritten as 
\begin{align}
\mathbf{J}\,\Delta \mathbf{x}
+
\mathbf{J}_{\alpha}\,\Delta \boldsymbol{\alpha}
\approx \mathbf{0},
\end{align}
Then, the change in the state can be expressed as 
\begin{align}
    \Delta \mathbf{x} = - \mathbf{J}^{-1} \mathbf{J}_{\alpha} \Delta \boldsymbol{\alpha}
\end{align}

The change in the admittance matrix for a $\ell-th$ line switching can be expressed as $|\Delta \boldsymbol{\alpha}| = |(1-z_\ell)\Delta \mathbf{y}|$, therefore using the first order approximation, the absolute change in the state due to switching of $\ell-th$ line can be approximated as
\begin{align}
    |\Delta \mathbf{x}| \approx |(1-z_\ell)\Delta \mathbf{y}|~|{\Xi}_l| 
    \label{eq:Foc_state}
\end{align}
\end{subequations}
where ${\Xi}_\ell $ as $\ell-$th column of $\mathbf{\Xi} = - \mathbf{J}^{-1} \mathbf{J}_{\alpha}$.
Below, we briefly explain how these sensitivities can be computed.

In \eqref{eq:Foc_state}, the matrix $\mathbf{J}$ is the usual power-flow Jacobian that is used during the power-flow solution in the Newton-Raphson scheme, whereas $\mathbf{J}_\alpha$ is the derivatives of the power injections with respect to the network admittance parameters. Our previous work in \cite{talkington2025differentiating} described a generalized way to compute first-order admittance sensitivities.

The power-flow Jacobian is given as 
$\mathbf{J} \triangleq \begin{bmatrix} J_{pv} & J_{pt} \\ J_{qv} & J_{qt} \end{bmatrix},$ containing the derivatives of the power injections with respect to the state $(|v|,\theta)$. Defining $\theta_{ij} = \theta_i - \theta_j$, the jacobian terms can be written as
\begin{subequations}
\label{eq:PFJac}
\begin{align}
(J_{pv})_{ii} &= 2|Y_{ii}| |v_i| \cos\phi_{ii} + \sum_{j\ne i} |Y_{ij}| |v_j| \cos(\theta_{ij}), \\
(J_{pv})_{ij} &= |Y_{ij}| |v_i| \cos(\theta_{ij}), \quad i\ne j, \\
(J_{qv})_{ii} &= -2|Y_{ii}| |v_i| \sin\phi_{ii} + \sum_{j\ne i} |Y_{ij}| |v_j| \sin(\theta_{ij}), \\
(J_{qv})_{ij} &= |Y_{ij}| |v_i| \sin(\theta_{ij}), \quad i\ne j,  \\
(J_{pt})_{ii} &= \sum_{j\ne i} \big(- |v_i| |Y_{ij}| |v_j| \sin(\theta_{ij})\big), \\
(J_{pt})_{ij} &= |Y_{ij}| |v_i| |v_j| \sin(\theta_{ij}), \quad i\ne j, \\
(J_{qt})_{ii} &= \sum_{j\ne i} \big(+ |v_i| |Y_{ij}| |v_j| \cos(\theta_{ij})\big), \\
(J_{qt})_{ij} &= -|Y_{ij}| |v_i| |v_j| \cos(\theta_{ij}), \quad i\ne j.
\end{align}
\end{subequations}

The admittance sensitivity matrix $\mathbf{J}_\alpha \triangleq \begin{bmatrix} \frac{\partial \mathbf{p}}{\partial \mathbf{g}} & \frac{\partial \mathbf{p}}{\partial \mathbf{b}} \\
\frac{\partial \mathbf{q}}{\partial \mathbf{g}} & \frac{\partial \mathbf{q}}{\partial \mathbf{b}} \end{bmatrix}$
consists of the derivatives of the power injections with respect to the line parameters. They are defined as follows.
\begin{subequations}
\begin{align}
    \frac{\partial p_i}{\partial g_{kj}} &= \begin{cases} 
        |v_k|^2 - |v_k| |v_j| \cos \theta_{kj} & \text{if } i=k \\
        |v_j|^2 - |v_k| |v_j| \cos \theta_{kj} & \text{if } i=j \\
        0 & \text{otherwise}
    \end{cases} \label{eq:dp_dg} \\
    \frac{\partial q_i}{\partial g_{kj}} &= \begin{cases} 
        -|v_k| |v_j| \sin \theta_{kj} & \text{if } i=k \\
        |v_k| |v_j| \sin \theta_{kj} & \text{if } i=j \\
        0 & \text{otherwise}
    \end{cases} \label{eq:dq_dg}\\
    \frac{\partial p_i}{\partial b_{kj}} &= \begin{cases} 
        -|v_k| |v_j| \sin \theta_{kj} & \text{if } i=k \\
        |v_k| |v_j| \sin \theta_{kj} & \text{if } i=j \\
        0 & \text{otherwise}
    \end{cases} \label{eq:dp_db} \\
    \frac{\partial q_i}{\partial b_{kj}} &= \begin{cases} 
        -|v_k|^2 + |v_k| |v_j| \cos \theta_{kj} & \text{if } i=k \\
        -|v_j|^2 + |v_k| |v_j| \cos \theta_{kj} & \text{if } i=j \\
        0 & \text{otherwise}
    \end{cases} \label{eq:dq_db}
\end{align}
\end{subequations}

Based on the above definitions, we define the penalty cost and constraints using the first-order admittance sensitivities for the proposed AC-informed DC-OTS formulation as follows.
\paragraph{First-order Admittance Sensitivity-based Repair Cost} 
To incorporate information about how line‑switching actions affect the network states, specifically, voltage magnitudes and phase angles, we introduce a penalty cost function based on the first‑order admittance sensitivities. These sensitivities quantify the linearized response of the AC network states to changes in line admittances, providing a tractable way to approximate the impact of switching operations.

By minimizing this penalty, the DC‑OTS formulation is steered away from topologies that would cause large deviations in the underlying AC states. In doing so, the model implicitly avoids switching decisions that are likely to lead to AC‑infeasible operating points. The resulting penalty‑augmented objective function is defined as:
\begin{align}
C_\text{repair}^\text{first} = 
\gamma_1 \sum_{\ell}
(1-z_\ell) ~ \|  \Xi_{\ell} \|_1
\label{eq:penalty}
\end{align}
where $\|\Xi_{\ell}\|$ refers to the norm-1 of  ${\Xi}_\ell$, $\gamma_1$ is a pre-defined weight, $z_\ell$ refers to binary variable for line switching.

The penalty term in \eqref{eq:penalty} is computed using the admittance sensitivities evaluated at the operating point obtained from the AC‑OPF under the base (pre‑switching) network topology. Because these sensitivities quantify how small perturbations in line admittances affect the AC network states, they provide a linearized measure of how disruptive a switching action is likely to be.
By minimizing this penalty, the DC‑OTS model is biased toward switching configurations that maintain the AC operating point close to the baseline, thereby reducing the likelihood of producing AC‑infeasible solutions.
\paragraph{First-order Admittance Sensitivity-based Constraint}
In addition to the penalty term in \eqref{eq:penalty}, which discourages topology choices likely to cause AC infeasibility, we also impose a constraint derived from the admittance sensitivities. This constraint is given by: \begin{align}
-{\epsilon}^{\text{lin}}_{\ell}
&\le (1-z_\ell)~\| \Xi_{\ell} \|_1 \le
{\epsilon}^{\text{lin}}_{\ell}, \quad \forall \ell, \label{eq:lin-bound}
\end{align}
Here, ${\epsilon}^{\text{lin}}_{\ell}$ specifies predefined bounds on the allowable changes in the system states, which are later derived in Section~\ref{sec:tolerances}. This constraint acts as a trust‑region condition, limiting how much the voltages can deviate due to a topology change.

\subsubsection{Second-order Admittance Sensitivity-based penalty and constraints}
The first‑order cost and constraint in \eqref{eq:penalty} and \eqref{eq:lin-bound} may be insufficient to capture the larger, nonlinear deviations in system states that can arise from topology changes. To better account for these nonlinear effects, we supplement the first‑order terms with a quadratic correction derived from the second‑order admittance sensitivities.

To obtain the second-order admittance sensitivities, we differentiate the 
power-flow equations in \eqref{eq:pf_eq} twice with respect to 
$\mathbf{(g,g)}$, $\mathbf{(b,b)}$, $\mathbf{(g,b)}$, and $\mathbf{(b,g)}$. 
The resulting differentiated systems are (see Appendix~\ref{sec:2ndorder}):
\begin{subequations}
\begin{align}
& \mathbf{J}\,\mathbf{S}_{gg}
+ \mathbf{H}_{xx}(\mathbf{S}_g,\mathbf{S}_g)
+ 2\,\mathbf{H}_{xg}\,\mathbf{S}_g
= \mathbf{0}, \label{eq:SOS1_S} \\
& \mathbf{J}\,\mathbf{S}_{bb}
+ \mathbf{H}_{xx}(\mathbf{S}_b,\mathbf{S}_b)
+ 2\,\mathbf{H}_{xb}\,\mathbf{S}_b
= \mathbf{0}, \label{eq:SOS2_S} \\
& \mathbf{J}\,\mathbf{S}_{gb}
+ \mathbf{H}_{xx}(\mathbf{S}_g,\mathbf{S}_b)
+ \mathbf{H}_{xg}\,\mathbf{S}_b
+ \mathbf{H}_{xb}\,\mathbf{S}_g
= \mathbf{0}, \label{eq:SOS3_S} \\
& \mathbf{J}\,\mathbf{S}_{bg}
+ \mathbf{H}_{xx}(\mathbf{S}_b,\mathbf{S}_g)
+ \mathbf{H}_{xb}\,\mathbf{S}_g
+ \mathbf{H}_{xg}\,\mathbf{S}_b
= \mathbf{0}. \label{eq:SOS4_S}
\end{align}
\end{subequations}
Here, $\mathbf{S}_g := \frac{\partial \mathbf{x}}{\partial \mathbf{g}}, 
\mathbf{S}_b := \frac{\partial \mathbf{x}}{\partial \mathbf{b}},$ are the first-order state sensitivities, and $\mathbf{S}_{gg} := \frac{\partial^2 \mathbf{x}}{\partial \mathbf{g}^2}, \quad
\mathbf{S}_{bb} := \frac{\partial^2 \mathbf{x}}{\partial \mathbf{b}^2}, \quad
\mathbf{S}_{gb} := \frac{\partial^2 \mathbf{x}}{\partial \mathbf{g}\partial \mathbf{b}}, \quad
\mathbf{S}_{bg} := \frac{\partial^2 \mathbf{x}}{\partial \mathbf{b}\partial \mathbf{g}},$ are the second-order state sensitivities. The tensor $\mathbf{H}_{xx}$ denotes the 
Hessian of $\mathbf{f}$ with respect to the state variables, defined as
$\mathbf{H}_{xx}(\mathbf{u},\mathbf{v}) 
= \sum_{i,j} 
\dfrac{\partial^2 \mathbf{f}}{\partial x_i\,\partial x_j}\, u_i v_j,$
and $\mathbf{H}_{xg}$ and $\mathbf{H}_{xb}$ denote the mixed second derivatives of 
$\mathbf{f}$ with respect to $(\mathbf{x},\mathbf{g})$ and $(\mathbf{x},\mathbf{b})$, 
respectively.

Equations \eqref{eq:SOS1_S} and \eqref{eq:SOS2_S} correspond to differentiation with 
respect to $\mathbf{g}$ and $\mathbf{b}$, while  \eqref{eq:SOS3_S} and \eqref{eq:SOS4_S} correspond to mixed $(\mathbf{g},\mathbf{b})$ 
and $(\mathbf{b},\mathbf{g})$ perturbations.
Solving \eqref{eq:SOS1_S}-\eqref{eq:SOS4_S} for the second-order sensitivities yields
\begin{subequations}
\begin{align}
& \mathbf{S}_{gg}
= -\mathbf{J}^{-1}
\left(
\mathbf{H}_{xx}(\mathbf{S}_g,\mathbf{S}_g)
+ 2\,\mathbf{H}_{xg}\,\mathbf{S}_g
\right), \label{eq:SOSGG_S} \\
& \mathbf{S}_{bb}
= -\mathbf{J}^{-1}
\left(
\mathbf{H}_{xx}(\mathbf{S}_b,\mathbf{S}_b)
+ 2\,\mathbf{H}_{xb}\,\mathbf{S}_b
\right), \label{eq:SOSBB_S} \\
& \mathbf{S}_{gb}
= -\mathbf{J}^{-1}
\left(
\mathbf{H}_{xx}(\mathbf{S}_g,\mathbf{S}_b)
+ \mathbf{H}_{xg}\,\mathbf{S}_b
+ \mathbf{H}_{xb}\,\mathbf{S}_g
\right), \label{eq:SOSGB_S} \\
& \mathbf{S}_{bg}
= -\mathbf{J}^{-1}
\left(
\mathbf{H}_{xx}(\mathbf{S}_b,\mathbf{S}_g)
+ \mathbf{H}_{xb}\,\mathbf{S}_g
+ \mathbf{H}_{xg}\,\mathbf{S}_b
\right). \label{eq:SOSBG_S}
\end{align}
\end{subequations}

The expressions in 
\eqref{eq:SOSGG_S}-\eqref{eq:SOSBG_S} represent the second-order admittance sensitivities, i.e., the second derivatives of the system state $\mathbf{x}$ with respect to the conductance $\mathbf{g}$, susceptance $\mathbf{b}$, and their mixed parameter combinations.

The hessian $\mathbf{H}_{xg}$ and $\mathbf{H}_{xb}$ consists of derivatives $\mathbf{p}, \mathbf{q}$ with respect to $|\mathbf{v}|, \boldsymbol{\theta}, \mathbf{g}, \mathbf{b}$. Let $\theta_{kj} = \theta_k - \theta_j $ it is given as
\begin{subequations}
\begin{align}
& \frac{\partial^2 p_k}{\partial |v_k| \partial g_{kj}}
= 2|v_k| - |v_j|\cos\theta_{kj}, 
\frac{\partial^2 p_k}{\partial \theta_k \partial g_{kj}} = |v_k| |v_j| \sin\theta_{kj}, \\
& \frac{\partial^2 q_k}{\partial |v_k| \partial g_{kj}}
= - |v_j| \sin\theta_{kj}, ~
\frac{\partial^2 q_k}{\partial \theta_k \partial g_{kj}} = - |v_k| |v_j| \cos\theta_{kj},\\
& \frac{\partial^2 p_k}{\partial |v_k| \partial b_{kj}}
= - |v_j| \sin\theta_{kj}, 
\frac{\partial^2 p_k}{\partial \theta_k \partial b_{kj}} = - |v_k| |v_j| \cos\theta_{kj}, \\
& \frac{\partial^2 q_k}{\partial |v_k| \partial b_{kj}}
= -2|v_k| + |v_j| \cos\theta_{kj}, 
\frac{\partial^2 q_k}{\partial \theta_k \partial b_{kj}} = |v_k| |v_j| \sin\theta_{kj}.
\end{align}
\end{subequations}

The Hessian $\mathbf{H}_{xx}$ contains the second derivatives of 
$\mathbf{p},\mathbf{q}$ with respect to $|\mathbf{v}|$ and 
$\boldsymbol{\theta}$. Let $\delta_{ij} = \theta_i - \theta_j - \phi_{ij}$. 

\begin{subequations}
\begin{align}
& \frac{\partial^2 p_i}{\partial |v_i|^2}
= 2|Y_{ii}|\cos\phi_{ii},
\frac{\partial^2 p_i}{\partial |v_i|\partial |v_j|}
= |Y_{ij}|\cos\delta_{ij},
\\
& \frac{\partial^2 p_i}{\partial |v_i|\partial \theta_i}
= \sum_{k\neq i} |v_k|\,|Y_{ik}|\sin\delta_{ik},
\frac{\partial^2 p_i}{\partial |v_i|\partial \theta_j}
= -|v_j|\,|Y_{ij}|\sin\delta_{ij},
\\
& \frac{\partial^2 p_i}{\partial |v_j|\partial \theta_i}
= |v_i|\,|Y_{ij}|\sin\delta_{ij},
\frac{\partial^2 p_i}{\partial |v_j|\partial \theta_j}
= -|v_i|\,|Y_{ij}|\sin\delta_{ij},
\\
& \frac{\partial^2 p_i}{\partial \theta_i^2}
= -|v_i|\sum_{k\neq i} |v_k|\,|Y_{ik}|\cos\delta_{ik},
\frac{\partial^2 p_i}{\partial \theta_i\partial \theta_j}
= |v_i||v_j|\,|Y_{ij}|\cos\delta_{ij},
\\
& \frac{\partial^2 q_i}{\partial |v_i|^2}
= -2|Y_{ii}|\sin\phi_{ii},
\frac{\partial^2 q_i}{\partial |v_i|\partial |v_j|}
= |Y_{ij}|\sin\delta_{ij},
\\
& \frac{\partial^2 q_i}{\partial |v_i|\partial \theta_i}
= -\sum_{k\neq i} |v_k|\,|Y_{ik}|\cos\delta_{ik},
\frac{\partial^2 q_i}{\partial |v_i|\partial \theta_j}
= |v_j|\,|Y_{ij}|\cos\delta_{ij},
\\
& \frac{\partial^2 q_i}{\partial |v_j|\partial \theta_i}
= -|v_i|\,|Y_{ij}|\cos\delta_{ij},
\frac{\partial^2 q_i}{\partial |v_j|\partial \theta_j}
= |v_i|\,|Y_{ij}|\cos\delta_{ij},
\\
& \frac{\partial^2 q_i}{\partial \theta_i^2}
= |v_i|\sum_{k\neq i} |v_k|\,|Y_{ik}|\sin\delta_{ik},
\frac{\partial^2 q_i}{\partial \theta_i\partial \theta_j}
= |v_i||v_j|\,|Y_{ij}|\sin\delta_{ij}.
\end{align}
\end{subequations}

The expressions for $\mathbf{J}$ is the same as in \eqref{eq:PFJac} and $\mathbf{S}_g, \mathbf{S}_b$ are elements of $\mathbf{\Xi} = - \mathbf{J}^{-1} \mathbf{J}_{\alpha}$.

Based on the above definitions, we define the Second-order repair cost and constraints as follows.
\paragraph{Second-order Admittance Sensitivity-based Repair Cost}
Similar to the first order cost function, we minimize penalty based on second-order sensitivities as 
\begin{align}
    C_{\text{repair}}^{\text{second}} = \gamma_2 \sum_\ell \tfrac12 (1-z_\ell)^2 \|\mathbb{S}_\ell\|_1
\end{align}
where ${\mathbb{S}}_\ell = {{S}_{gg}}_\ell + {{S}_{bb}}_\ell + {{S}_{gb}}_\ell + {{S}_{bg}}_\ell
$ is defined as sum of $\ell-$th columns of all the Hessians.
The symbols ${{S}_{gg}}_\ell,~ {{S}_{bb}}_\ell,~ {{S}_{bg}}_\ell,~{{S}_{gb}}_\ell$ refer to $\ell-$th column of ${\mathbf{S}_{gg}},~ {\mathbf{S}_{bb}},~ {\mathbf{S}_{bg}},~ {\mathbf{S}_{gb}}$. The symbol $\gamma_2$ denotes a pre-defined weight.

The quadratic term $ \tfrac12 (1-z_\ell)^2 \|\mathbb{S}_\ell\|_1$ accounts for the curvature of the AC equations in addition to the first-order correction. Large curvature means the linear correction is inaccurate, and the trajectory due to a change in the topology may deviate from the original state by a large margin. Penalizing this term avoids topologies that push the solution outside the local AC-validity region reducing the AC infeasibility.
\paragraph{Second-order admittance sensitivity-based constraint}
Similar to the first-order case, we also impose a constraint using the second-order sensitivities as 
\begin{align}
-{\epsilon}^{\text{quad}}_\ell
&\le \tfrac12(1-z_\ell)^2 \|\mathbb{S}_\ell\|_1 \le
{\epsilon}^{\text{quad}}_\ell, \quad \forall \ell. \label{eq:quad-bound}
\end{align}
where, ${\epsilon}^{\text{quad}}_\ell$ is pre-defined bound are later computed in Section~\ref{sec:tolerances}. 

\subsection{Final AC-informed DC-OTS problem (AIDC-OTS)}
Finally, we present the new AC-informed DC-OTS formulation, which includes the constraints and the penalty costs based on the admittance sensitivities. For the sake of completeness, the full formulation is given as 
\begin{subequations}
\begin{align}
& \begin{aligned}
\min_{P_g,\bm{\theta},\bm{f},\bm{z}} ~ & \sum_{g\in\mathcal{G}} C_g(P_g) +  \gamma_1 \sum_{\ell}
(1-z_\ell) ~\|\Xi_l \|_1 + \\ & \gamma_2 \sum_\ell \tfrac12 (1-z_\ell)^2 \|{S}_\ell\|_1
\end{aligned} \label{eq:aiddc-ots-obj}\\
\text{s.t.}\quad 
& \sum_{g\in\mathcal{G}_i} P_g - P_{d,i} = \sum_{\ell\in\delta(i)} \sigma_{i\ell} f_\ell,\quad \forall i\in\mathcal{N}, \label{eq:balance}\\
& f_\ell - b_\ell(\theta_i-\theta_j) \le M_\ell(1-z_\ell),\ \forall \ell\in\mathcal{L}, \label{eq:bigM1}\\
& f_\ell - b_\ell(\theta_i-\theta_j) \ge -M_\ell(1-z_\ell),\ \forall \ell\in\mathcal{L}, \label{eq:bigM2}\\
& -\bar f_\ell z_\ell \le f_\ell \le \bar f_\ell z_\ell,\quad \forall \ell\in\mathcal{L}, \label{eq:flowlimit}\\
& P_g^{\min} \le P_g \le P_g^{\max},\quad \forall g\in\mathcal{G}, \label{eq:genlimit}\\
& z_\ell\in\{0,1\},\quad \forall \ell\in\mathcal{L}. \label{eq:binary_ots} \\
& -{\epsilon}^{\text{lin}}_\ell
\le (1-z_\ell)~\|\Xi_\ell \|_1 \le
{\epsilon}^{\text{lin}}_\ell, \quad \forall \ell,\\
& -{\epsilon}^{\text{quad}}_\ell
\le \tfrac12(1-z_\ell)^2 \|\mathbb{S}_\ell\|_1 \le
{\epsilon}^{\text{quad}}_\ell,\quad \forall \ell \label{eq:quad_const}.
\end{align}
\label{eq:AIDC-OTS}
\end{subequations}
As it can be observed, the AIDC-OTS problem in \eqref{eq:AIDC-OTS} is a mixed integer quadratic constrained quadratic program (MIQCQP) due to the quadratic objective and constraints, which may have higher computational complexity compared to the MILP. In the following, we show two different approaches, such as Linear and SOCP reformulations, to achieve MILP and MISOCP  formulations.

\subsection{Linear and SOCP Reformulation}\label{sec:realizations}
The quadratic term in the objective and in the constraint in \eqref{eq:AIDC-OTS} makes the AIDC-OTS problem an MIQCQP, which may be slower, especially on large systems. In this section, we introduce two different approximations to make it MISOCP and MILP, which are solver-friendly formulations.

\subsubsection{Linear reformulation}
The quadratic terms with binaries, $(1-z_\ell)^2$, can be approximated as follows. 
\begin{align}
    (1-z_\ell)^2 = (1-z_\ell)
\end{align}
This transforms the quadratic constraint and objective in \eqref{eq:aiddc-ots-obj} and \eqref{eq:quad_const} into linear terms.
\subsubsection{SOCP reformulation}
Let $t_\ell=1-z_\ell\in[0,1]$ and $v_\ell=\|\mathbb{S}_\ell\| t_\ell$, then the constraint in \eqref{eq:quad_const} can be re-written as
\begin{align}
    \norm{0.5\,v_\ell}_1 \le \epsilon_{\text{quad}},\quad v_\ell=\|\mathbb{S}_\ell\|_1 t_\ell, \ t_\ell=1-z_\ell.
\end{align}

Similarly, the objective can be approximated as 
\begin{align}
    C_{\text{repair}}^{\text{second, socp}} = \norm{0.5\,v_\ell}_1
\end{align}

With the above reformulations, both the linear and SOCP reformulations are convex and typically faster than MIQCQP.

\subsection{Systematic Construction of Line-Specific Tolerances}
\label{sec:tolerances}
In this section, we present an approach to compute the line-specific tolerances $\varepsilon^{\text{lin}}_\ell$ and 
$\varepsilon^{\text{quad}}_\ell$ used to bound the residuals of AC-informed sensitivity constraints in our OTS formulation. The construction follows a Taylor-style approximation of the AC power-flow mapping with respect to line switching and scales naturally with the physical sensitivity of each line.

\subsubsection{Taylor-Style Error Budgeting}
Combining the first and second order terms as per the Taylor series expansion of the power-flow equations in \eqref{eq:pf_eq}, the change in the state due to a change in the series admittance can be expressed as
\begin{subequations}
\begin{align}
\Delta \mathbf{x}_\ell \;\approx\; \Xi_\ell \cdot \|\Delta y_\ell\| \;+\;
\frac{1}{2}\, \mathbb{S}_\ell \cdot \|\Delta y_\ell\|^2 \;+\; \mathcal{O}\bigl(\|\Delta y_\ell\|^3\bigr),
\label{eq:taylor}
\end{align}
where the vectors $\Xi_\ell$ and $\mathbb{S}_\ell$ summarize the $\ell-th$ column of first- and
second-order sensitivities at the operating point. Taking norms on both sides
and upper-bounding the higher-order terms yields the standard error-budgeting
relations
\begin{align}
\bigl\| \text{1st-order tolerance for line }\ell \bigr\|
&\;\lesssim\; \|\Xi_\ell\| \, \|\Delta y_\ell\|,
\label{eq:bound-lin}\\[2pt]
\bigl\| \text{2nd-order tolerance for line }\ell \bigr\|
&\;\lesssim\; \frac{1}{2}\, \|\mathbb{S}_\ell\| \, \|\Delta y_\ell\|^2.
\label{eq:bound-quad}
\end{align}
In our implementation and experiments, we use the $\ell_1$-norm to match the
penalty and constraint structure, i.e., $\|\cdot\|\equiv\|\cdot\|_1$.
\subsubsection{Tolerances from Sensitivity Magnitudes}
We translate \eqref{eq:bound-lin}–\eqref{eq:bound-quad} into
\emph{line-specific tolerances} via two pre-defined factors
$\alpha>0$ and $\beta>0$:
\label{eq:eps-linquad}
\begin{align}
\varepsilon^{\text{lin}}_\ell
&\;\triangleq\; \alpha_1 \,\|\Xi_\ell\|_p \;\|y_\ell\|_p,
\label{eq:eps-lin}\\
\varepsilon^{\text{quad}}_\ell
&\;\triangleq\; \alpha_2 \,\frac{1}{2}\,\|\mathbb{S}_\ell\|_p \;\|y_\ell\|^{\,2},
\label{eq:eps-quad}
\end{align}
where $p\in\{1,2\}$ is the chosen norm index (we use $p=1$ by default).
The scalars $(\alpha_1,\alpha_2)$ provide mild conservativeness (we used
$\alpha_1=1$, $\alpha_2=1$ across all cases) and can be fixed globally
to ensure reproducibility. For numerical robustness, we apply small floors
(and optional caps):
\begin{align}
\varepsilon^{\text{lin}}_\ell & \leftarrow \min\Bigl\{\max\bigl(\varepsilon^{\text{lin}}_\ell,\;\underline{\varepsilon}^{\text{lin}}\bigr),\;\overline{\varepsilon}^{\text{lin}}\Bigr\}, \\
\varepsilon^{\text{quad}}_\ell & \leftarrow \min\Bigl\{\max\bigl(\varepsilon^{\text{quad}}_\ell,\;\underline{\varepsilon}^{\text{quad}}\bigr),\;\overline{\varepsilon}^{\text{quad}}\Bigr\},
\end{align}
\end{subequations}
with small floors $\underline{\varepsilon}^{\text{lin}}=\underline{\varepsilon}^{\text{quad}}=10^{-6}$
and caps $\overline{\varepsilon}^{\text{lin}}=\overline{\varepsilon}^{\text{quad}}=+\infty$
unless stated otherwise.
The tolerances in \eqref{eq:eps-lin}–\eqref{eq:eps-quad} adapt automatically to network size and operating point via norm of the first- and second-order admittance sensitivities. 

\section{Numerical Results}
To assess the performance of the proposed AC-Informed DC-OTS formulation, we conduct a comprehensive evaluation across a diverse set of publicly available test networks drawn from the PGLib-OPF benchmark library \cite{babaeinejadsarookolaee2019power}. These include standard IEEE test networks ranging from small (\texttt{case14\_ieee}) to large-scale systems (\texttt{case1888\_rte}, \texttt{case2383\_wp}). This variety enables a quantitative comparison of scalability, AC feasibility, and computational efficiency across all OTS formulations considered in this work.

\subsection{Algorithm Implementation}
The implementation detail is presented in Algorithm~\ref{alg:acinf-ots}. We start by solving the AC-OPF on the base topology, which is used to obtain the initial state and the first- and second-order admittance sensitivities. Then, different formulations, i.e., original DC-OTS, proposed AIDC-OTS, LPAC-OTS, and QC-OTS, are solved and evaluated under AC-OPF.
\begin{algorithm}[!htbp]
\caption{AC-Informed OTS}
\label{alg:acinf-ots}
\begin{algorithmic}[1]
\STATE Solve the AC-OPF on the base topology. Record the AC-OPF objective for comparison 
\STATE Using the AC-OPF solution, obtain the state $(\mathbf{x}_0)$ which is used to compute the first- and second-order admittance sensitivities in $\bm{\Xi},\mathbb{S}$ via the procedure in Section~\ref{subsec:AIDOTS:theory}.
\STATE Build DC-OTS constraints and costs as well as the first and second order constraints using admittance sensitivities. 
\STATE Solve the SOCP or linear formulation. If solved, report the model cost and lines opened. 
\STATE If solved, re-evaluate the optimized topology using the AC-OPF, and report the re-evaluated AC-OPF cost.
\end{algorithmic}
\end{algorithm}

We benchmark the proposed AC-informed DC-OTS against the original DC-OTS model, and two existing reformulations of the AC-OTS problem, which are LPAC-OTS and QC-OTS. 
All the implementations are done in Julia; AC-OPF, DC-OTS, LPAC-OTS, and QC-OTS are implemented using the PowerModels.jl library \cite{coffrin2018powermodels}. 

\begin{table*}[!htbp]
\centering
\caption{Performance Comparison of the proposed AIDC-OTS with DC, LPAC, and QC-OTS Across Various Test Cases: Lines Opened, Optimized Cost, AC-OPF Evaluation Cost, and Solve Time}
\label{tab:multi_case_ots_time}
\renewcommand{\arraystretch}{1.15}
\setlength{\tabcolsep}{5pt}

\begin{tabular}{
>{\raggedright\arraybackslash}p{2.3cm}
>{\raggedright\arraybackslash}p{2.7cm}
>{\raggedright\arraybackslash\footnotesize}p{5.5cm}
r r c}
\toprule
\textbf{Test Case} &
\textbf{Method} &
\textbf{Lines Opened (indices)} &
\textbf{Model Cost [\$]} &
\textbf{Evaluated on AC-OPF [\$]} &
\textbf{Time [s]} \\
\midrule
\multirow{5}{*}{\texttt{case14\_ieee}}
 & Baseline AC-OPF        & --                & 2178.08 & 2178.08 & 0.01 \\
 & DC-OTS                & \{3, 5, 9, 11, 14, 19, 20\}
                         & 2051.52 & Inf. & 0.04 \\
 & AIDC-OTS (proposed)    & none              & 2051.52 & 2178.08 & 0.08 \\
 & LPAC-OTS              & none              & 2167.60 & 2178.08 & 0.14 \\
 & QC-OTS                & \{19\}            & 2175.78 & 2178.14 & 0.43 \\
\midrule

\multirow{5}{*}{\texttt{case30\_ieee}}
 & Baseline AC-OPF        & --                & -- & 8208.51 & 0.03 \\
 & DC-OTS                &
 \{3, 5, 12, 13, 16, 19, 20, 24, 29, 30, 35, 37, 41\}
                         & 5639.29 & Inf & 0.02 \\
 & AIDC-OTS (proposed)    & \{3, 5\}           & 5639.31 & Inf & 0.11 \\
 & LPAC-OTS              & \{3, 14, 24, 26, 29, 35\}
                         & 7424.12 & 7585.63 & 1.73 \\
 & QC-OTS                & \{33\}             & 6662.05 & 8207.43 & 0.65 \\
\midrule

\multirow{5}{*}{\texttt{case39\_epri}}
 & Baseline AC-OPF        & --                & -- & 138415.56 & 0.04 \\
 & DC-OTS                & \{4, 6, 7, 9, 15, 21, 44\}
                         & 135860.94 & 138453.86 & 0.06 \\
 & AIDC-OTS (proposed)    & \{4, 6, 30\}
                         & 135860.94 & 137743.59 & 0.13 \\
 & LPAC-OTS              & \{6, 30\}
                         & 137603.74 & 137717.34 & 0.51 \\
 & QC-OTS                & none
                         & 137664.01 & 138415.56 & 0.46 \\
\midrule

\multirow{5}{*}{\texttt{case118\_ieee}}
 & Baseline AC-OPF        & -- & -- & 97213.60 & 3.0 \\
 & DC-OTS                &
 \{2, 14-15, 17-18, 20, 26, 42, 44, 49, 55, 58, 62, 64, 68, 70-71, 73, 79, 93, 98, 114-115, 119-121, 130, 140, 146, 149, 154, 156, 166, 168-169, 171, 178, 182\}
                         & 93026.72 & Inf. & 1.18 \\
 & AIDC-OTS (proposed)    & \{63, 65, 106, 166\}
                         & 93031.58 & 96729.05 & 2.06 \\
 & LPAC-OTS              & none
                         & 96060.61 & 97213.60 & 1800$^{\dagger}$ \\
 & QC-OTS                & \{30, 44, 63, 65, 78, 80, 106\}
                         & 96373.93 & 96648.36 & 31.0 \\
\midrule

\multirow{5}{*}{\texttt{case179\_goc}}
 & Baseline AC-OPF        & -- & -- & 754266.41 & 0.24 \\
 & DC-OTS                & *86 lines opened
                         & 751169.81 & 755304.05 & 0.48 \\
 & AIDC-OTS (proposed)    & \{19, 22, 162, 167, 240-241, 251-252,263\}
                         & 751193.97 & 753967.53 & 0.48 \\
 & LPAC-OTS              & none
                         & 753443.82 & 754266.41 & 1800$^{\dagger}$ \\
 & QC-OTS                & \{8, 26, 35-36, 52, 74-76, 81, 84, 89, 107-108, 118, 164, 172, 175, 194, 234-235, 239, 250, 263\}
                         & 753106.60 & {754140.42} & 125.26 \\
\midrule

\multirow{5}{*}{\texttt{case200\_active}}
 & Baseline AC-OPF        & -- & -- & 27557.57 & 0.13 \\
 & DC-OTS                & \{115, 116, 138, 219, 221, 222, 223, 225, 226, 244, 245\}
                         & 27479.64 & 27557.57 & 0.07 \\
 & AIDC-OTS (proposed)    & none
                         & 24659.12 & 27557.57 & 0.11 \\
 & LPAC-OTS              & \{18, 23, 44, 85, 102, 116, 138, 139, 142, 165, 169, 219, 222-223, 225-226, 244-245\}
                         & 27557.41 & 27559.60 & 289.05 \\
 & QC-OTS                & \{23, 115, 116, 119, 138, 142, 169, 219, 221, 222, 223, 225, 226, 244, 245\}
                         & 27556.82 & 27557.72 & 803.83 \\
\midrule

\multirow{5}{*}{\texttt{case300\_ieee}}
 & Baseline AC-OPF        & -- & -- & 565219.97 & 0.52 \\
 & DC-OTS                & *97 lines opened 
                         & 505721.10 & Inf. & 1800$^{\dagger}$ \\
 & AIDC-OTS (proposed)    & none
                         & 517532.45 & 565219.97 & 626.30 \\
 & LPAC-OTS              & none
                         & 552150.20 & 565219.97 & 1800$^{\dagger}$ \\
 & QC-OTS                & ns.
                         & ns. & ns. & 1800$^{\dagger}$ \\
\midrule
\multirow{5}{*}{\texttt{case500\_goc}}
 & Baseline AC-OPF        & -- & -- & 454758.74 & 0.61 \\
 & DC-OTS                & 
 *151 lines opened
                         & 439882.47 & Inf. & 2.64 \\
 & AIDC-OTS (proposed)    & \{427\}
                         & 412603.07 & 454758.74 & 0.55 \\
 & LPAC-OTS              & ns.
                         & ns. & ns. & 1800$^{\dagger}$ \\
 & QC-OTS                & ns.
                         & ns. & ns. & 1800$^{\dagger}$ \\
\midrule
\multirow{5}{*}{\texttt{case1888\_rte}}
 & Baseline AC-OPF        & -- & --  & 1402530.82 & 5.5 \\
 & DC-OTS                & *88 lines opened
                         & 1352871.75 & Inf. & 3.42 \\
 & AIDC-OTS (proposed)    & 30 lines opened 
                         & 1334895.35 & {1375862.57} & 3.21 \\
 & LPAC-OTS              & ns.
                         & ns. & ns. & 1800$^{\dagger}$ \\
 & QC-OTS                & ns.
                         & ns. & ns. & 1800$^{\dagger}$ \\
\midrule
\multirow{5}{*}{\texttt{case2383\_wp}}
 & Baseline AC-OPF        & -- & --  & 1868191.58 & 3.71 \\
 & DC-OTS                & *487 lines opened 
                         & 1768481.04 & Inf. & 1800$^{\dagger}$  \\
 & AIDC-OTS (proposed)    & *79 lines opened 
                         & 1789156.50 &   {1865543.13} & 1800$^{\dagger}$  \\
 & LPAC-OTS              & ns.
                         & ns. & ns. & 1800$^{\dagger}$  \\
 & QC-OTS                & ns.
                         & ns. & ns. & 1800$^{\dagger}$  \\
\midrule

\end{tabular}
\vspace{-1mm}
\begin{tablenotes}
    \item $^*$ For brevity, the number of lines opened are presented instead of all the indices. $^{\dagger}$ Maximum time limit reached.
\end{tablenotes}
\end{table*}
\subsection{Results}
We present simulation results across several benchmark test cases, summarized in Table~\ref{tab:multi_case_ots_time}. For each case study, we report the base-topology AC-OPF cost, the optimized objective value of the OTS model (model cost), the number of opened lines, the AC-OPF cost evaluated on the optimized topology, and the total solution time.  For all the simulations, we consider a maximum time limit of 1800s. For the cases, when it was not solved within this time, it is reported as Not solved within the time limit (ns). ``Inf'' means the model was found to be infeasible. The case-wise results are detailed below.

\subsubsection{\texttt{case14\_ieee}}
For \texttt{case14\_ieee}, the base AC-OPF cost is \$2178.08. The DC-OTS solution yields a model cost of \$2051.52 with 7 lines opened; however, this topology is AC-infeasible when re-evaluated on AC-OPF. The proposed AIDC-OTS does not open any lines and achieves the same cost while remaining AC-feasible. LPAC-OTS also chooses not to open any lines with a comparable cost. QC-OTS opens one line and yields a similar objective value.

\subsubsection{\texttt{case30\_ieee}}
For this system, the base AC-OPF cost is \$8208.51. Both DC-OTS and AIDC-OTS produce topologies that become AC-infeasible upon re-evaluation. The best-performing model is LPAC-OTS, which achieves an AC-feasible cost of \$7583.63, representing a 7.61\% improvement.

\subsubsection{\texttt{case39\_epri}}
DC-OTS opens 7 lines and results in an AC re-evaluated cost of \$138{,}453.86, exceeding the base AC-OPF cost and illustrating that DC-OTS can increase costs under AC evaluation. AIDC-OTS attains \$137{,}743.59, a 0.5\% reduction relative to the base case and an improvement over DC-OTS and QC-OTS. LPAC-OTS yields the lowest cost of \$137{,}717.34, slightly outperforming AIDC-OTS.
\subsubsection{\texttt{case118\_ieee}}
In this case, DC-OTS opens 37 lines, but the resulting topology is AC-infeasible. AIDC-OTS identifies an AC-feasible topology with cost slightly below the base AC-OPF and better than LPAC-OTS, though QC-OTS performs marginally better. In terms of computational time, LPAC-OTS reaches the 1800~s limit, QC-OTS completes in 31~s, and AIDC-OTS requires only 2.1~s.
\subsubsection{\texttt{case179\_goc}}
The base AC-OPF cost is \$754{,}266.41. DC-OTS opens 86 lines and produces an AC cost of \$755,304.05. The proposed AIDC-OTS achieves the best cost of \$753,967.53, outperforming others.
\subsubsection{\texttt{case200\_active}}
For this test case, all OTS models except QC-OTS yield AC-feasible topologies, but none improve upon the base AC-OPF cost. LPAC-OTS again does not converge within the maximum time limit.

\subsubsection{\texttt{case300\_ieee}}
DC-OTS opens 97 lines, resulting in an AC-infeasible topology. Both AIDC-OTS and LPAC-OTS choose not to open any lines. LPAC-OTS reaches the time limit, and QC-OTS does not solve within the allotted time.

\subsubsection{\texttt{case500\_goc}}
DC-OTS opens 151 lines, which leads to AC infeasibility. AIDC-OTS opens only one line and preserves AC feasibility, though without cost improvement. LPAC-OTS and QC-OTS fail to converge. 

\subsubsection{\texttt{case1888\_rte}}
The base AC-OPF cost is \$1{,}402{,}530.82. DC-OTS opens 88 lines and produces an AC-infeasible topology. AIDC-OTS opens 30 lines and yields an AC-feasible cost of \$1{,}375{,}862.57, corresponding to a 1.9\% improvement. LPAC-OTS and QC-OTS do not converge within the time limit.

\subsubsection{\texttt{case2383\_wp}}
This system exhibits behavior similar to \texttt{case1888\_rte}. The base AC-OPF cost is \$1{,}868{,}191.58. DC-OTS opens 487 lines, resulting in AC infeasibility. AIDC-OTS opens 79 lines and achieves an AC-feasible cost of \$1{,}865{,}543.13, yielding a 0.15\% improvement. LPAC-OTS and QC-OTS do not solve within the allotted time.

\subsection{Sensitivity Analysis with Penalty Parameters}
The objective function for AIDC-OTS contains weight parameters $\gamma_1, \gamma_2$ that need to be set for simulations. In the following, we perform sensitivity analysis with different values of $\gamma_1, \gamma_2$. We present the sensitivity analysis for two different test networks \texttt{case118\_ieee} and \texttt{case1888\_rte}. For simplicity, we show the sensitivity analysis for the case when both $\gamma_1$ and $\gamma_2$ are changed simultaneously.

\begin{table}[!htbp]
\centering
\caption{Impact of $\gamma_1, \gamma_2$ for \texttt{case118\_ieee}}
\vspace{-1em}
\label{tab:gamma_cost_case118}
\footnotesize
\setlength{\tabcolsep}{4pt}
\begin{tabular}{c | p{1.6cm} | p{5.6cm}}
\hline
$\bm \gamma_1, \bm \gamma_2$ & \bf Evaluated on AC-OPF (Model Cost) &  \bf Lines Opened \\
\hline
0.01 & 98249.03 (93026.73) & [6, 16, 24, 34, 66, 67, 78, 94, 118, 149, 170, 174] \\
\hline
0.05 & 97075.15 (93026.73) & [46, 63, 65, 78, 79, 80, 91, 106, 122, 123, 170, 174] \\
\hline
0.10 & 97605.15 (93026.73) & [3, 16, 70, 73, 77, 78, 83, 90, 100, 121, 123, 149, 170, 174] \\
\hline
0.50 & 98037.19 (93029.13) & [12, 29, 34, 66, 67, 77, 78, 118, 122, 170, 174] \\
\hline
1.00 & 97638.96 (93026.73) & [26, 46, 61, 65, 76, 77, 81, 82, 88, 90, 100, 101, 123, 128, 146, 154, 160, 168, 174] \\
\hline
5.00 & Inf (93027.04) & [49, 56, 66, 67, 68, 79, 94, 110, 125, 148, 165, 166, 168, 169, 171] \\
\hline
10.0 & Inf (93026.74) & [44, 57, 58, 61, 77, 78, 79, 80, 82, 101, 117, 122, 154, 166, 168, 169, 171, 178, 182, 186] \\
\hline
\end{tabular}
\vspace{-1em}
\end{table}
The sensitivity analysis for the \texttt{case118\_ieee} is shown in Table~\ref{tab:gamma_cost_case118}. We vary the weights from 0.01 to 10, and report the model cost (optimized cost), the cost obtained by evaluating the optimized topology with AC-OPF, and the lines opened. As can be seen, the model cost does not vary with the weights significantly and is almost the same in each case and close to the original DC-OTS model cost in Table~\ref{tab:multi_case_ots_time}. It is interesting to note that when $\gamma_1, \gamma_2$ is varied from 0.01 to 1, the model is AC feasible, whereas for the values higher than 1, the optimized topologies are AC-infeasible. This example demonstrates that it is important to choose these weights wisely. 

We also present sensitivity analysis for a larger test case, \texttt{case1888\_rte}, in Table~\ref{tab:gamma_cost1888}.  
Again, we compare the model cost, the cost evaluated on AC-OPF, and the lines opened. In this case, for the $\gamma_1, \gamma_2$ of 0.1 to 10, we observe AC-infeasibility, whereas for values higher than 10, eg, 100 and 1000, we obtain AC-feasible solutions. Another observation is that with higher weight, the number of lines opened reduces, which might be due to higher penalty cost in \eqref{eq:aiddc-ots-obj}.
\begin{table}[!t]
\centering
\caption{Impact of $\gamma_1, \gamma_2$ for \texttt{case1888\_rte}}
\vspace{-1em}
\label{tab:gamma_cost1888}
\footnotesize
\setlength{\tabcolsep}{4pt}
\begin{tabular}{c | p{1.6cm} | p{5.6cm}}
\hline
$\bm \gamma, \bm \beta$ & \bf Evaluated on AC-OPF (Model Cost) &  \bf Lines Opened \\
\hline
0.1 & inf (1334895.35) & [9, 31, 103-104, 117, 309-310, 321, 336-338, 351, 360, 372, 402, 408, 482, 538, 717, 723-734, 764, 835, 843, 963, 1055, 1074, 1172, 1198, 1238, 1294, 1346, 1422, 1428, 1518, 1531-1532, 1559, 1560, 1577, 1598, 1641, 1679, 1708, 1729, 1749, 1751, 1855, 1939, 2046, 2093-2094, 2517] \\
\hline
1.00 & inf (1334895.35) & [[103, 104, 117, 336, 337, 338, 360, 372, 402, 408, 538, 557, 563, 587, 734, 764, 843, 963, 1055, 1074, 1156, 1172, 1198, 1238, 1346, 1428, 1437, 1518, 1531, 1532, 1560, 1577, 1641, 1651, 1708, 1719, 1814, 1939, 2046, 2093, 2094]] \\
\hline
10.00 & Inf (1334895.35) &  [37, 103, 104, 336, 337, 338, 372, 402, 408, 538, 587, 720, 723, 734, 764, 1055, 1156, 1172, 1198, 1238, 1346, 1428, 1518, 1531, 1532, 1577, 1708, 1814, 1939, 2046, 2093, 2094]\\
\hline
100.0 & 1375857.27 (1334895.35) & [32, 103, 104, 337, 338, 538, 764, 1055, 1156, 1172, 1198, 1238, 1346, 1428, 1518, 1531, 1532, 1577, 1939, 2046, 2093, 2094]
\\
1000 &  1375857.27 (1334895.35) & 103, 104, 538, 764, 1055, 1172, 1238, 1346, 1428, 1518, 1531, 1577, 1939, 2046, 2093, 2094]
\\
\hline
\end{tabular}
\vspace{-1em}
\end{table}
\subsection{Sensitivity Analysis with Constraints Conservativeness}
We also perform sensitivity analysis with the parameters that are used in constraints, i.e., tolerances in eqs.~\eqref{eq:eps-lin}-\eqref{eq:eps-quad}. We vary the weights $\alpha_1, \alpha_2$ and simulate the AIDC-OTS scheme. The results for \texttt{case118\_ieee} and \texttt{case1888\_rte} are presented in Tables~\ref{tab:alpha_cost118} and \ref{tab:alpha_cost1888}, respectively.

As shown in Table~\ref{tab:alpha_cost118}, for the case of \texttt{case118\_ieee}, the model cost, the AC-OPF evaluated cost, and the number of lines change with $\alpha_1, \alpha_2$. We achieve the minimum cost for $\alpha_1, \alpha_2 \geq 1$. For the $\alpha_1, \alpha_2 < 1$, the costs are higher, and no lines are opened.
In \texttt{case1888\_rte}, we see that for all values of $\alpha_1, \alpha_2$, there are always some lines to be opened. The model cost remains the same for all cases, whereas the minimum AC-OPF evaluated is found at $\alpha_1, \alpha_2 \geq 1$.
\begin{table}[!t]
\centering
\caption{Impact of $\alpha_1, \alpha_2$ for \texttt{case118\_ieee}}
\vspace{-1em}
\label{tab:alpha_cost118}
\footnotesize
\setlength{\tabcolsep}{4pt}
\begin{tabular}{c | p{1.6cm} | p{5cm}}
\hline
$\bm \alpha_1, \bm \alpha_2 $ & \bf Evaluated on AC-OPF (Model Cost) &  \bf Lines Opened \\
\hline
0.01 & 97213.61 (93132.67) & no lines opened \\
\hline
0.05 & 97213.61 (93132.67) & no lines opened \\
\hline
0.10 & 97213.61 (93132.67)  & no lines opened  \\
\hline
0.50 & 97213.61 (93132.67)  & no lines opened  \\
\hline
1.00 & 97075.15 (93026.72) & [46, 63, 65, 78, 79, 80, 91, 106, 122, 123, 170, 174] \\
\hline
5.00 & 97075.15 (93026.72) & [46, 63, 65, 78, 79, 80, 91, 106, 122, 123, 170, 174] \\
\hline
10.0 & 97075.15 (93026.72) & [46, 63, 65, 78, 79, 80, 91, 106, 122, 123, 170, 174] \\
\hline
\end{tabular}
\vspace{-1em}
\end{table}
\begin{table}[!t]
\centering
\caption{Impact of $\alpha_1, \alpha_2$ for \texttt{case1888\_rte}}
\vspace{-1em}
\label{tab:alpha_cost1888}
\footnotesize
\setlength{\tabcolsep}{4pt}
\begin{tabular}{c | p{1.6cm}  | p{5cm}}
\hline
$\bm \alpha_1, \bm \alpha_2 $ & \bf Evaluated on AC-OPF (Model Cost) &  \bf Lines Opened \\
\hline
0.001 & 1375861.43 (1334895.35) & [103, 104, 337, 338, 538, 764, 843, 1055, 1172, 1198, 1238, 1346, 1428, 1518, 1531, 1532, 1577, 1939, 2046, 2093, 2094] \\
\hline
0.01 & 1375861.43 (1334895.35) & [103, 104, 337, 338, 538, 764, 843, 1055, 1172, 1198, 1238, 1346, 1428, 1518, 1531, 1532, 1577, 1939, 2046, 2093, 2094] \\ 
\hline
0.1 & 1375861.43 (1334895.35) & [103, 104, 337, 338, 538, 764, 843, 1055, 1172, 1198, 1238, 1346, 1428, 1518, 1531, 1532, 1577, 1939, 2046, 2093, 2094] \\  
\hline
1 & 1375857.27 (1334895.35) & [32, 103, 104, 337, 338, 538, 764, 1055, 1156, 1172, 1198, 1238, 1346, 1428, 1518, 1531, 1532, 1577, 1939, 2046, 2093, 2094] \\
\hline
10 & 1375857.27 (1334895.35) & [32, 103, 104, 337, 338, 538, 764, 1055, 1156, 1172, 1198, 1238, 1346, 1428, 1518, 1531, 1532, 1577, 1939, 2046, 2093, 2094] \\
\hline
\end{tabular}
\vspace{-2em}
\end{table}
\subsection{Discussion and Limitations}
In the following section, we provide some discussion on the presented results and the key limitations of the work. 
\begin{itemize}[leftmargin=*]
    \item \textbf{AC-feasibility and Optimality:} Based on the simulated cases, we observed that the proposed AIDC-OTS scheme achieved AC-feasibility as well as lower cost than the original DC-OTS formulation. This demonstrates the advantages of the proposed scheme. However, there are some cases when the cost achieved by the AIDC-OTS were slighly higher compared to the LPAC-OTS and QC-OTS schemes, e.g., \texttt{case39\_epri} and \texttt{case118\_ieee}. Although it should be noted that the costs are lower compared to the original DC-OTS formulation. It should also be noted that there could be cases when AIDC-OTS fails in ensuring feasibility, for example, for \texttt{case30\_ieee}, both the DC-OTS and AIDC-OTS lead to AC-infeasibility when re-evaluated on AC-OPF. This may happen when the admittance sensitivities are too small in magnitude.
    \item \textbf{Computational tractability:} As observed in the results, the proposed AIDC-OTS is higly tractable algorithm, thanks to the DC-OTS linear structure. For all cases (except \texttt{case2383\_wp}), the scheme converged before the maximum time limit of 1800s. For most cases, except \texttt{case300\_ieee}, it takes similar or even less time than the original DC-OTS, which is in the order of 0.1 - 4s. On the other hand, the existing AC-OTS approximations, such as LPAC-OTS and QC-OTS, are always computationally heavy and non-converging for bigger test cases, such as the LPAC-OTS scheme could not be solved for \texttt{case500\_goc}, \texttt{case1888\_rte} and \texttt{case2383\_wp}, whereas QC-OTS could not be solved for \texttt{case300\_ieee}, \texttt{case500\_goc}, \texttt{case1888\_rte} and \texttt{case2383\_wp}.
    \item \textbf{Sensitivity with respect to the weight parameters:}
    Based on the sensitivity analysis with respect to the weight parameters, it was observed that there are cases when AIDC-OTS may lead to AC-infeasible solutions; however, in most cases it is AC-feasible with similar costs. So, when the results are AC-feasible, the cost does not change a lot. Therefore, it is necessary to choose wisely these weight parameters. As the AIDC-OTS formulation is computationally tractable, this sensitivity analysis can be carried out efficiently. 
\end{itemize}

\section{Conclusion}
The DC-OTS formulation, although computationally efficient, is usually AC-infeasible and sub-optimal when the optimized topology is re-evaluated on the AC-OPF. This leads to unreliable switching decisions that can not be implemented on real networks.

To tackle this issue, this work proposed a new AC-informed DC-OTS scheme to obtain an AC-feasible solution with a realistic number of lines to be opened. This was achieved by the first- and second-order Admittance sensitivities-based optimization model by augmenting the DC-OTS formulation with AC repair cost and constraints. These repair costs and the constraints guide the AIDC-OTS model to achieve an AC-feasible topology and relatively optimal costs. Furthermore, we presented a linearized and SOCP equivalent formulation of the AIDC-OTS model, which are solver friendly. 

The proposed scheme was evaluated on a large number of publicly available test networks. The results show that the proposed AIDC-OTS scheme gives AC-feasible solutions and a lower cost compared to the existing DC-OTS model. Also, the proposed scheme is faster compared to the other AC-OTS approximations, i.e., LPAC and QC-OTS models. The sensitivity analysis on the weight factors demonstrated that there might be cases when the proposed AIDC-OTS might be AC-infeasible if the weights are not chosen wisely. Future work will focus on extending the scheme for other DC-power flow-based MILP formulations. 

\appendix
\vspace{-0.5em}
\subsection{Second-order derivatives}
\label{sec:2ndorder}
Below shows brief derivation for \eqref{eq:SOS1_S}-\eqref{eq:SOS4_S}.
Differentiating, $f(x, \alpha)$ twice with respect to $\alpha$ gives
\footnotesize
\begin{subequations}
\begin{align}
\frac{\partial}{\partial \alpha_k}
\left(
\frac{\partial f}{\partial x_i}
\frac{\partial x_i}{\partial \alpha_j}
+
\frac{\partial f}{\partial \alpha_j}
\right)
= & 0 \\
\Rightarrow \begin{aligned}
\frac{\partial^2 f}{\partial x_i \partial x_s} 
\frac{\partial x_s}{\partial \alpha_k}
\frac{\partial x_i}{\partial \alpha_j} & 
+
\frac{\partial f}{\partial x_i}
\frac{\partial^2 x_i}{\partial \alpha_j \partial \alpha_k}
+ \\
&
\frac{\partial^2 f}{\partial x_i \partial \alpha_k}
\frac{\partial x_i}{\partial \alpha_j}
+
\frac{\partial^2 f}{\partial \alpha_j \partial \alpha_k}
= 0.
\end{aligned}\\
\Rightarrow  \begin{aligned}
\frac{\partial f}{\partial x_i}
\frac{\partial^2 x_i}{\partial \alpha_j \partial \alpha_k}
&=
-
\Bigg(
\frac{\partial^2 f}{\partial x_i \partial x_s}
\frac{\partial x_i}{\partial \alpha_j}
\frac{\partial x_s}{\partial \alpha_k}
+
\frac{\partial^2 f}{\partial x_i \partial \alpha_k}
\frac{\partial x_i}{\partial \alpha_j}
+ \\
& \frac{\partial^2 f}{\partial x_i \partial \alpha_j}
\frac{\partial x_i}{\partial \alpha_k}
+
\frac{\partial^2 f}{\partial \alpha_j \partial \alpha_k}
\Bigg).
\end{aligned}\\
\Rightarrow  \begin{aligned}
\frac{\partial^2 x_i}{\partial \alpha_j \partial \alpha_k}
= & 
-
\left(\mathbf{J}^{-1}\right)_{i\ell}
\Bigg(
\frac{\partial^2 f}{\partial x_i \partial x_s}
\frac{\partial x_i}{\partial \alpha_j}
\frac{\partial x_s}{\partial \alpha_k}
+ \\
& \frac{\partial^2 f}{\partial x_i \partial \alpha_j}
\frac{\partial x_i}{\partial \alpha_k}
+
\frac{\partial^2 f}{\partial x_i \partial \alpha_k}
\frac{\partial x_i}{\partial \alpha_j}
+
\frac{\partial^2 f}{\partial \alpha_j \partial \alpha_k}
\Bigg)
\end{aligned}
\end{align}
\begin{align}
\mathbf{J}\,\mathbf{S}_{\alpha\alpha}
=
-
\Big[
\mathbf{H}_{xx}\!\left(\mathbf{S}_\alpha,\mathbf{S}_\alpha\right)
+
\mathbf{H}_{x\alpha}\mathbf{S}_\alpha
+
\left(\mathbf{H}_{x\alpha}\mathbf{S}_\alpha\right)^{\!\top}
+
\mathbf{H}_{\alpha \alpha}
\Big]
\end{align}
\end{subequations}
\vspace{-2em}
\bibliographystyle{IEEEtran}
\bibliography{refs.bib}

\end{document}